\documentclass[11pt]{article}
\usepackage{xspace}
\usepackage{graphicx}
\usepackage{amsmath}
\usepackage{amssymb}
\usepackage{color}
\usepackage{caption}
\usepackage{subcaption}
\definecolor{Red}{rgb}{1,0,0}
\definecolor{Green}{rgb}{0,1,0}
\definecolor{Blue}{rgb}{0,0,1}
\definecolor{Black}{rgb}{0,0,0}

\def\beq{\begin{equation}}
\def\eeq#1{\label{#1}\end{equation}}
\def\eeqn{\end{equation}}

\def\beqa{\begin{eqnarray}}
\def\eeqa#1{\label{#1}\end{eqnarray}}
\def\eeqan{\end{eqnarray}}

\let\bar=\overbar

\def\Dslash{\not{\hbox{\kern-4pt $D$}}}
\def\dslash{\not{\hbox{\kern-2pt $\del$}}}

\def\msb{{\bar{\ssstyle M \kern -1pt S}}}

\def\Title#1{\begin{center} {\Large {\bf #1} } \end{center}}

\begin{document}

\Title{Results from T2K}

\bigskip\bigskip


\begin{raggedright}  

Martin David Haigh\index{Haigh, M. D.}, {\it University of Warwick}\\

\begin{center}\emph{On behalf of the T2K Collaboration.}\end{center}
\bigskip
\end{raggedright}

{\small
\begin{flushleft}
\emph{To appear in the proceedings of the Interplay between Particle and Astroparticle Physics workshop, 18 -- 22 August, 2014, held at Queen Mary University of London, UK.}
\end{flushleft}
}

\section{Introduction}

The Tokai to Kamioka (T2K) experiment uses a beam of muon neutrinos,
produced at the J-PARC facility on the east coast of Japan, to study
neutrino oscillations driven by the $\Delta m_{\textrm{atm}}^{2}$ mass
splitting. A suite of near detectors located 280~m from the secondary
beam source samples the unoscillated beam, and the Super-Kamiokande
water Cherenkov detector samples the beam at a baseline of 295~km, and
at a point $2.5^\circ$ off the beam axis, giving a narrow-band beam
centred around 600~MeV. Analyses of the oscillation channels $\nu_\mu
\to \nu_e$ and $\nu_\mu \to \nu_\mu$ allow measurements to be made of
$\theta_{13}$, $\theta_{23}$ and $\Delta m^2_{\textrm{atm}}$, and,
ultimately, for weak constraints to be placed on the CP-violating
phase $\delta_{CP}$.

In addition to these analyses, T2K has made world-leading neutrino
cross-section measurements in the sub-GeV energy range, utilising both
the near and far detectors. The present work will discuss both the
most recent measurements of the oscillation parameters, and these
cross section analyses.

\section{Neutrino oscillations}

Observations of neutrino flavour change conclusively favour a model
whereby the neutrinos have non-zero mass, and the mass eigenstates of
the neutrino are not identical with the weak (flavour)
eigenstates. Instead they are related by a unitary transformation,
known as the Pontecorvo-Maki-Nakagawa-Sakata (PMNS) matrix
$U_{\textrm{PMNS}}$:
\begin{equation}
\begin{pmatrix}
\nu_e \\ \nu_\mu \\\nu_\tau \\
\end{pmatrix}
=U_{\textrm{PMNS}}
\begin{pmatrix}
\nu_1 \\ \nu_2 \\ \nu_3 \\
\end{pmatrix}.
\end{equation}

Since a neutrino produced in a specific weak eigenstate $\nu_\alpha$
is in a superposition of mass, and therefore energy, eigenstates, then
after propagating some distance the different phases between the
mass states will produce a superposition of flavour states. The
neutrino may therefore be observed in a different flavour state
$\nu_\beta$. The probability for a neutrino produced in a flavour
eigenstate $\nu_\alpha$ to be observed later in a flavour eigenstate
$\nu_\beta$, depends on the neutrino energy $E$, the propagation
distance $L$, the mass splittings between the eigenstates, $\Delta
m^2_{ij}\equiv m^2_j-m^2_i$, as well as on the elements $U_{\alpha i}$
of the mixing matrix, as
\begin{align}
P(\nu_\alpha \to \nu_\beta)=&\delta_{\alpha \beta} \nonumber \\
-&4\sum\limits_i \sum\limits_{j \leq i}
\Re \left(U^*_{\alpha i} U_{\beta i} U_{\alpha j} U^*_{\beta j}\right)
\sin^2 \left(\Delta m_{ij}^2 \frac{L}{4E}\right) \\
+&2\sum\limits_i \sum\limits_{j \leq i} 
\Im \left(U^*_{\alpha i} U_{\beta i} U_{\alpha j} U^*_{\beta j}\right)
\sin \left(\Delta m_{ij}^2 \frac{L}{2E}\right). \nonumber
\end{align}

This unitary transformation may be written as the product of three
two-dimensional rotations with angles
$(\theta_{12},\theta_{13},\theta_{23})$, with the addition of a
complex phase $\delta_{CP}$, and CP-violation will be present unless
$\delta_{CP}$ is equal to 0 or $\pi$. The additional degrees of
freedom in the unitary transformation correspond to phases which are
of no consequence for oscillation physics. Specifically, the matrix
may be factorised thus:
\begin{equation}
U_{\text{PMNS}}=
\begin{pmatrix}
1 & 0 & 0 \\
0 & c_{23} & s_{23} \\
0 & -s_{23} & c_{23} \\
\end{pmatrix}
\begin{pmatrix}
c_{13} & 0 & s_{13}e^{-i\delta_{CP}} \\
0 & 1 & 0 \\
-s_{13}e^{+i\delta_{CP}} & 0 & c_{13} \\
\end{pmatrix}
\begin{pmatrix}
c_{12} & s_{12} & 0 \\
-s_{12} & c_{12} & 0 \\
0 & 0 & 1 \\
\end{pmatrix},
\end{equation}
where $(s_{ij},c_{ij})$ correspond respectively to the sin and cosine
of the angle $\theta_{ij}$. For the case of T2K, it is informative to
write the oscillation probabilities in terms of these mixing angles,
neglecting terms containing $\Delta m^2_{12}$, since at T2K energies
and baselines, $\Delta m_{12}^2 \frac{L}{E}\ll 1$. This leads to the
simplified expressions
\begin{align}
P(\nu_\mu \to \nu_\mu)\approx & 
1-\sin^2(2\theta_{23})\sin^2 \left( \frac{\Delta m^2_{23}L}{4E} \right) \\
P(\nu_\mu \to \nu_e)\approx & 
\sin^2 (\theta_{23}) \sin^2 (2\theta_{13}) 
\sin^2 \left( \frac{\Delta m^2_{23}L}{4E} \right).
\end{align}
It can be seen that when considering only terms in $\Delta m^2_{23}$,
the $\nu_\mu$ survival probability depends only on the angle
$\theta_{23}$, and the $\nu_e$ appearance probability on $\theta_{13}$
and $\theta_{23}$. Dependencies on $\theta_{12}$, $\delta_{CP}$, and
extra terms due to the presence of matter, appear only at higher
orders.

\section{The T2K Experiment}
T2K is a long-baseline neutrino oscillation experiment based in
Japan~\cite{Abe:2011ks}. A beam of muon neutrinos is produced at the
J-PARC facility using 30~GeV protons from the Main Ring (MR)
accelerator, in a conventional fashion. The protons are extracted and
impacted onto a target to produce a secondary beam containing mostly
$\pi$ and $K$, which are then focused using a series of three magnetic
horns, whose polarity can be changed to focus positive or negative
hadrons, producing a dominantly $\nu_\mu$ or $\bar{\nu}_\mu$ beam
respectively. Neutrinos are produced in the decays of hadrons; the
dominant process for a $\nu_\mu$ beam is $(\pi^+ \to \mu^+ +
\nu_\mu)$. Due to the presence of other decays and wrong-sign hadrons
in the beam, an admixture of electron neutrinos, and wrong-sign muon
neutrinos, is also present.

The T2K experiment is novel in that the main detectors are placed
approximately $2.5^\circ$ from the axis of the neutrino beam, giving a
narrow-band beam with neutrino energies tightly focused around
600~MeV, the energy of the first oscillation maximum at the far
detector. It also removes most of the high-energy tail of the beam,
reducing the background from high-energy deep inelastic and resonant
$\pi$ production events, which can be mis-reconstructed as
quasi-elastic events at a lower energy.

The far detector for T2K is the Super-Kamiokande water Cherenkov
detector, at Mozumi Mine near Kamioka, on the west coast of
Japan. Super-K is 295~km from J-PARC, and has a rock overburden of
1~km (2.7~km water equivalent), reducing non-beam backgrounds to a
negligible level. It has a 50~kt total water mass, and 22.5~kt
fiducial mass. The detector is cylindrical and split into an inner
detector (ID) and outer veto region (OD), both instrumented with PMTs.
Neutrino interactions are identified by observing the Cherenkov
radiation from the resultant charged lepton; muons and electrons are
distinguished by the different ring shapes produced, with muons
producing a clean ring, and electrons a fuzzy ring due to showering.

A near detector complex at the J-PARC site, 280~m from the target, is
used to constrain the properties of the unoscillated neutrino beam. It
consists of an on-axis component (INGRID) and a detector at the same
off-axis angle as Super-K (ND280). The INGRID consists of thirteen
$1\times1\times1~m^3$ iron-plastic scintillator sampling calorimeter
modules arranged in a plus configuration centred on the beam axis,
with two additional off-diagonal modules. Its purpose is to measure
the overall intensity and direction of the beam.

The ND280 detector is a general-purpose detector comprising multiple
components. At the upstream end there is a module (the P0D) optimised
for reconstruction of $\pi^0$ events. Downstream of this is a tracker
region, consisting of two fine-grained scintillator detectors (FGDs)
which make up the target mass, sandwiched between three gas
TPCs. Surrounding the inner detectors are electromagnetic sampling
calorimeters (ECALs), and a 0.2~T magnet to allow reconstruction of
the momentum and sign of charged particles in the TPCs. The yoke of
the magnet is instrumented with side muon range detectors (SMRDs),
which also serve as a cosmic trigger.

The results presented in this paper are based on data taken in the
period 2010--2013 (Runs 1--4). This period corresponds to a total of
$6.6\times 10^{20}$ 30~GeV protons on target. T2K is presently
gathering $\bar{\nu}_\mu$ data for analysis.

\section{Flux constraint from beam simulation and ND280}
\label{Section:ND280}
The nominal beam flux, including admixtures, is modelled using a Monte
Carlo, which uses GEANT3 to track hadrons through the target and
horns, and internally implements the relatively simple propagation and
decay of particles in the beam pipe. Results of the NA61 hadron
production experiment at CERN are used to constrain hadron production
in the target.

\begin{figure}[!ht]
  \begin{center}
    \begin{subfigure}[b]{0.3\textwidth}
      \includegraphics[width=\textwidth]{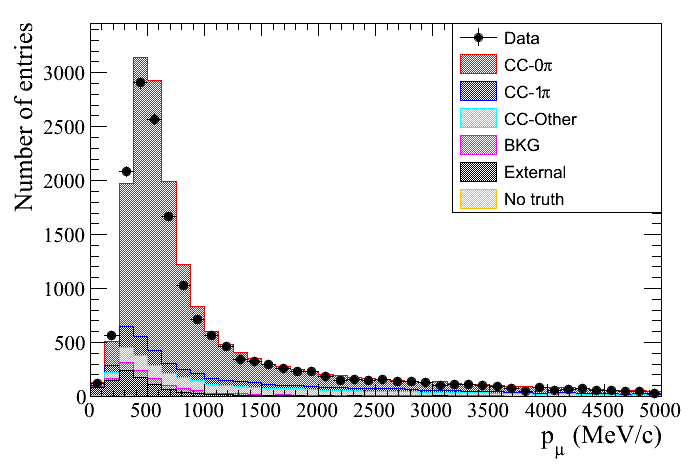}
      \caption{$\nu_\mu$CC-$0\pi$}
    \end{subfigure}
    \begin{subfigure}[b]{0.3\textwidth}
      \includegraphics[width=\textwidth]{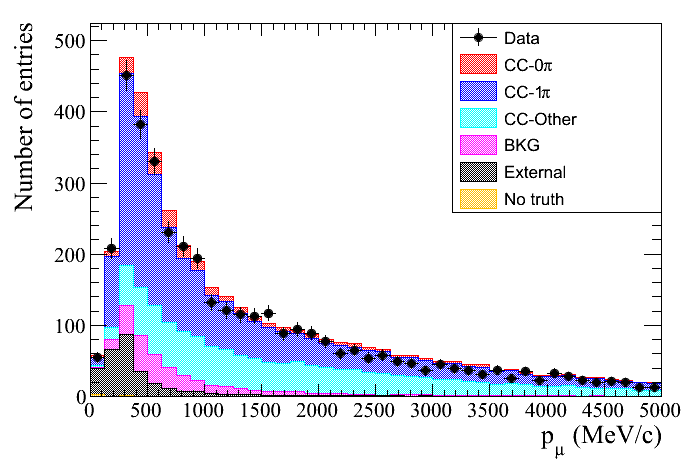}
      \caption{$\nu_\mu$CC-$1\pi$}
    \end{subfigure}
    \begin{subfigure}[b]{0.3\textwidth}
      \includegraphics[width=\textwidth]{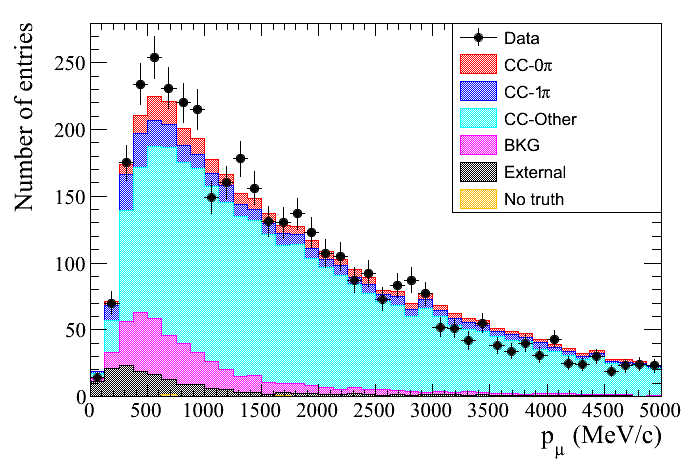}
      \caption{$\nu_\mu$CC-other}
    \end{subfigure}
    
    \caption{$\nu_\mu$-CC event spectra in the ND280, for data and nominal Monte Carlo}
\label{fig:ND280Spectra}
\end{center}
\end{figure}

Results from the ND280 are used to tune the nominal Monte Carlo. The
analysis used for this tuning is based on identifying and classifying
charged-current (CC) interactions of muon neutrinos in the most
upstream FGD of the detector. $\nu_\mu$-CC events are tagged by
looking for a vertex in FGD1, where the highest-energy negative track
in the TPC immediately downstream has energy deposition consistent
with a muon. These events are then classified as CC-$0\pi$,
CC-$1\pi^+$ or CC-other, based on the pions identified in the final
state. Charged pions are tagged by pion-like TPC tracks, or, for
positive pions, Michel electrons in FGD1. Neutral pions are tagged
using electron-like tracks in the TPC. Any event containing multiple
pions, or a $\pi^0$ or $\pi^-$, falls into the CC-other
category. Figure \ref{fig:ND280Spectra} shows the momentum spectra for
all events in these categories, both in data and nominal Monte Carlo.

\begin{figure}[!ht]
  \begin{center}
    \begin{subfigure}[b]{0.4\textwidth}
      \includegraphics[width=\textwidth]{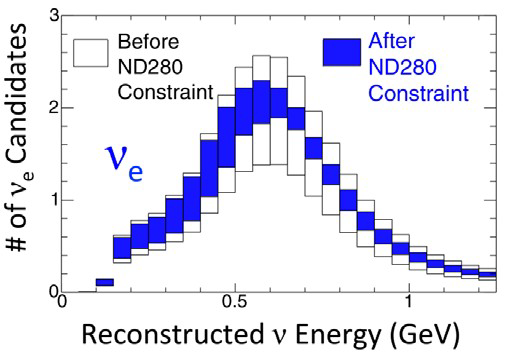}
      \caption{}
    \end{subfigure}
    \begin{subfigure}[b]{0.4\textwidth}
      \includegraphics[width=\textwidth]{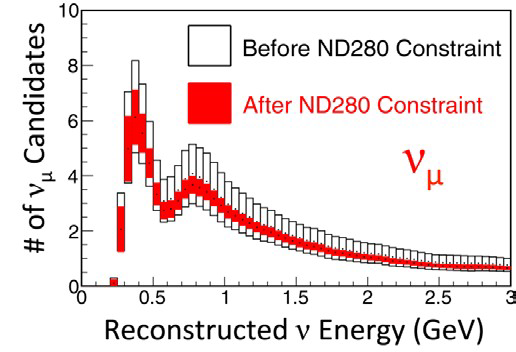}
      \caption{}
    \end{subfigure}
    
    \caption{Predicted event spectra at Super-K, before and after constraints from fitting to ND280 data. Shown for $\nu_e$ (a) and $\nu_\mu$ (b).}
\label{fig:eventRatePredictions}
\end{center}
\end{figure}

The beam Monte Carlo parameters and cross-section models, with prior
external constraints, are fitted to the three ND280 samples, to
calculate the expected event rate at Super-K. The resulting improvement
in event rate predictions can be seen in Figure
\ref{fig:eventRatePredictions}.

\section{Muon neutrino disappearance analysis}

Muon neutrino disappearance at T2K is observed via a deficit in the
number of $\nu_\mu$ events seen at Super-K relative to that predicted
in the absence of oscillations~\cite{Abe:2014ugx}. This deficit is
expected to be energy-dependent, and enables measurements of
$\theta_{23}$ and $\Delta m^2_{23}(\Delta m^2_{31})$ for
normal(inverted) hierarchy (i.e. the sign of the mass splitting
$\Delta m^2_{23}$), hereafter referred to as NH or IH.

A cuts-based analysis is used to obtain an event sample enriched with
CCQE events ($\nu_\mu+n \to \mu^- + p$). The proton is usually below
the Cherenkov threshold and therefore invisible. The analysis demands
a single ring, with muon-like PID, vertex and whole event in the
fiducial volume, low activity in the OD, visible energy $>$ 30~MeV,
reconstructed $p_\mu$ $>$ 200~MeV, and $\leq 1$ decay electron. The
event must be in time with the beam. For selected events, the neutrino
energy is reconstructed assuming that the underlying neutrino
interaction is quasi-elastic.

Oscillation parameters are estimated from this selection using an
unbinned maximum-likelihood fit to the events. A full three-flavour
fit, including matter effects, is performed. The oscillation
parameters $\Delta m^2_{23}(\Delta m^2_{31})$ for NH(IH),
$\delta_{CP}$ and $\sin^2 \theta_{23}$ are allowed to float freely in
the fit. $\sin^2 \theta_{13}$, $\sin^2 \theta_{12}$ and $\Delta
m^2_{21}$ are fitted with external constraints from reactor and
solar+KamLAND data. 45 systematic parameters are included in the fit,
from cross section uncertainties common with ND280 and specific to
Super-K, Super-K detector effects, final state (FSI) and secondary
(SI) interactions, and the aforementioned constrained oscillation
parameters. The systematic errors from these sources are broken down
in Table \ref{Table:NuMuSys}.

\begin{table}[!th]
  \begin{center}
    \begin{tabular}{l | c}
      \hline
      Source of uncertainty (number of parameters) & $\delta n^\textrm{exp}_\textrm{SK} / n^\textrm{exp}_\textrm{SK}$ \\
      \hline
      ND280-independent cross section (11) & 4.9\% \\
      Flux \& ND280-common cross section (23) & 2.7\% \\
      SK detector \& FSI+SI systematics (7) & 5.6\% \\
      $\sin^2 \theta_{13}$, $\sin^2 \theta_{12}$, $\Delta m^2_{21}$, $\delta_{CP}$ (4) & 0.2\% \\
      \hline
      Total (45) & 8.1\% \\
      \hline
      \end{tabular}
  \end{center}
      \caption{Effect of 1$\sigma$ systematic parameter variation on
        the number of 1-ring $\mu$-like events, computed for
        oscillations with $\sin^2 \theta_{23}=0.500$ and $|\Delta
        m^2_{32}|=2.40\times 10^{-3}~eV^2 / c^4$.}
      \label{Table:NuMuSys}
\end{table}

\begin{figure}[!ht]
  \begin{center}
      \includegraphics[width=0.4\textwidth]{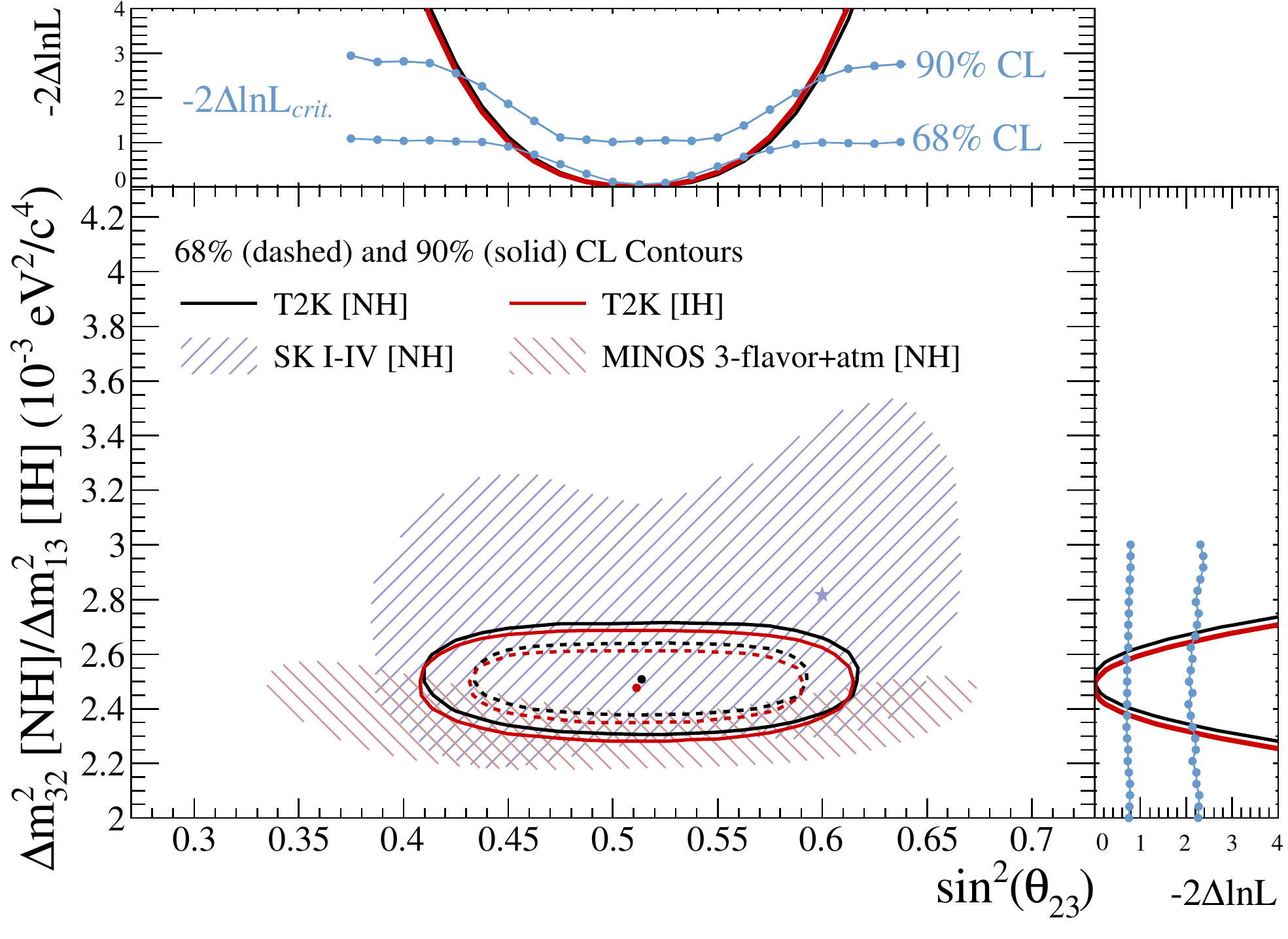}
    \caption{Oscillation best fit point and confidence regions for the
      T2K $\nu_\mu$ analysis, with MINOS and Super-K results for
      comparison. External constraints on other oscillation parameters
      are: ${\sin^2 \theta_{13}=0.0251\pm 0.0035}$, ${\sin^2
        \theta_{12}=0.312 \pm 0.016}$, ${\Delta m^2_{21}=(7.50 \pm
        0.20) \times 10^{-5}}$~eV${}^2/c^4$. $\delta_{CP}$ is
      unconstrained and a uniform Earth matter density of
      2.6~g/cm${}^3$ is used.}
\label{fig:nuMuResult}
\end{center}
\end{figure}

Figure \ref{fig:nuMuResult} shows the central values and 2D confidence
regions for $\theta_{23}$ and $\Delta m^2_{23}(\Delta m^2_{31})$ for
NH(IH), along with 90\% C.L. limits for MINOS and Super-K for
comparison. Confidence regions are calculated using the
Feldman-Cousins technique. The best fit values and 1D 68\%
C.L. intervals are $\sin^2 \theta_{23}=0.514^{+0.055}_{-0.056}(0.511\pm 0.055)$ and $\Delta m_{23}^2=2.51\pm 0.10 \times 10^{-3}$~eV${}^2$/c${}^4$($\Delta m_{31}^2=2.48\pm 0.10\times 10^{-3}$~eV${}^2$/c${}^4$) for NH(IH). The error budget is
dominated by statistical errors.

\section{Electron neutrino appearance analysis}

To measure electron neutrino appearance, the number of electron
neutrino events, and their spectral shape, are compared to those
expected in the absence of oscillations~\cite{Abe:2013hdq}. Background
events come from the intrinsic $\nu_e$ component of the beam, and also
from mis-reconstructed $\nu_\mu$ events, in particular NC-$\pi^0$
($\nu_\mu + p \to \nu_\mu + p + \pi^0$), where one $\gamma$ from the
pion decay is missed. The analysis allows measurement of $\sin^2 2
\theta_{13}$.

The $\nu_e$ selection at Super-K is made using similar preselection
cuts as for the $\nu_\mu$ analysis, on event timing, vertex location
and event containment, visible energy and OD activity. As with the
$\nu_\mu$ analysis, a CCQE-enriched sample is obtained by selecting
events containing a single electron-like ring, with reconstructed
momentum greater than 100~MeV/c. Events with a reconstructed neutrino
energy (assuming CCQE kinematics) greater than 1250~MeV are also
rejected, since these are dominantly due to the intrinsic $\nu_e$ beam
component. To remove $\pi^0$ background, an algorithm measuring the
maximum likelihood of the observed PMT hit pattern, assuming
$\nu_e$-CCQE or NC-$\pi^0$, is used; events are accepted or rejected
based on the reconstructed $\pi^0$ mass and the likelihood ratio
between the $\pi^0$ and electron hypothesis best fits. These cuts lead
to an expected 21.6 selected events for $\sin^2 2\theta_{13}=0.1$ and
$\delta_{CP}=0$; the expected number for $\theta_{13}=0$ is 4.92. In
actuality, 28 candidate events were observed. The total systematic
error on the number of observed events is 8.8\% for $\sin^2
2\theta_{13}=0.1$;this is dominated by neutrino cross-section
uncertainties.

\begin{figure}[!ht]
  \begin{center}
      \includegraphics[width=0.4\textwidth]{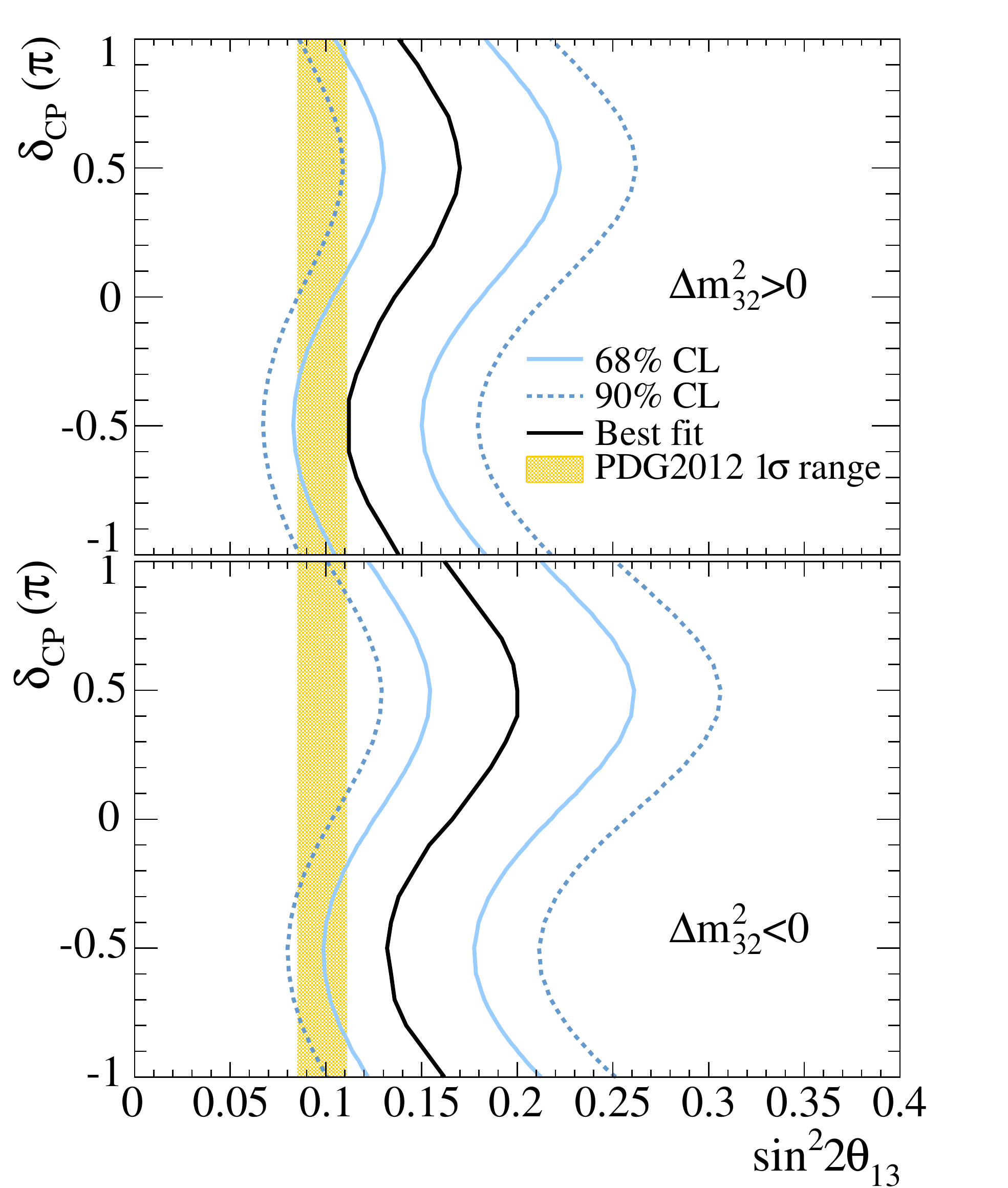}
    \caption{Oscillation best fit point and confidence regions for the
      T2K $\nu_e$ analysis.}
\label{fig:nuEResult}
\end{center}
\end{figure}

A binned extended maximum-likelihood fit is used to extract
oscillation parameters from the data. The likelihood function contains
terms for the signal shape, signal normalisation, T2K systematics and
external oscillation constraints. Events are parameterised using the
reconstructed electron momentum and angle; however, parameterising
events using reconstructed neutrino energy gives compatible best-fit
points and near-identical confidence regions. Other oscillation
parameter inputs are as follows: $\sin^2 \theta_{12}=0.306$, $\Delta
m^2_{21}=7.6 \times 10^{-5}$~eV${}^2$, $\sin^2 \theta_{23}=0.5$, $|\Delta
m^2_{32}|=2.4 \times 10^{-3}$~eV${}^2$, $\delta_{CP}=0$. For the NH(IH) the
best-fit values, and 68\% confidence level bounds, are $\sin^2 2
\theta_{13}=9.140^{+0.038}_{-0.032}(0.170^{+0.045}_{-0.037})$. The
significance of non-zero $\theta_{13}$ is found to be $7.3\sigma$; the
same result is obtained with either a delta log-likelihod method or
toy Monte Carlo. This significance was obtained using the values for
$\sin^2 \theta_{23}$ and $\delta_{CP}$ as above; however the
significance remains above $7.0\sigma$ for any values of these
parameters consistent with their uncertainties. The $\sin^2 2
\theta_{13}$ confidence limits are shown as a function of
$\delta_{CP}$ in Figure \ref{fig:nuEResult}.

\section{Cross section measurements}

The T2K experiment is making world leading neutrino cross section
measurements in the $\sim 1$~GeV range. Only results and outlines of
the analyses will be presented here; the reader is directed to the
references given for detailed descriptions.

\subsection{$\nu_e$ CC-inclusive cross section on Carbon}

The CC-inclusive cross section of $\nu_e$ on Carbon has been
measured in the ND280 detector~\cite{Abe:2014agb}. The signal for this
process is any event containing an electron in the final state, and
the principal background is $\gamma$s from $\pi^0$ decay.

Events are selected which have a negative electron-like track (by
$\text{d}E/\text{d}x$) in TPC2, with a vertex in FGD1. Events are
subject to a veto on activity upstream of FGD1. The $\pi^0$ background
is reduced by demanding that events containing a positive track do not
have an invariant mass compatible with a $\pi^0$ when the track is
matched with the electron candidate.

\begin{figure}[!ht]
  \begin{center}
    \begin{subfigure}[b]{0.4\textwidth}
      \includegraphics[width=\textwidth]{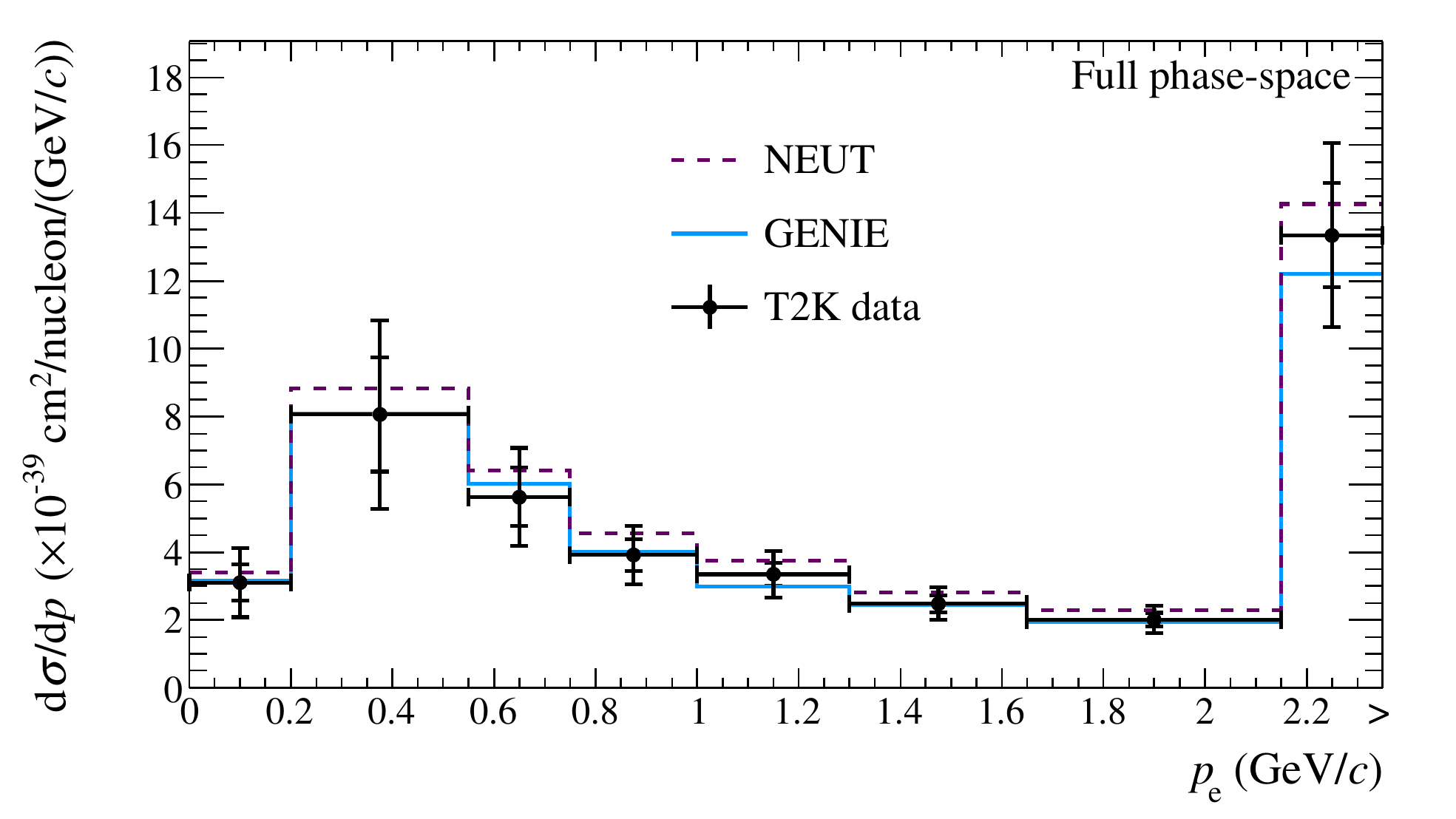}
      \caption{}
    \end{subfigure}
    \begin{subfigure}[b]{0.4\textwidth}
      \includegraphics[width=\textwidth]{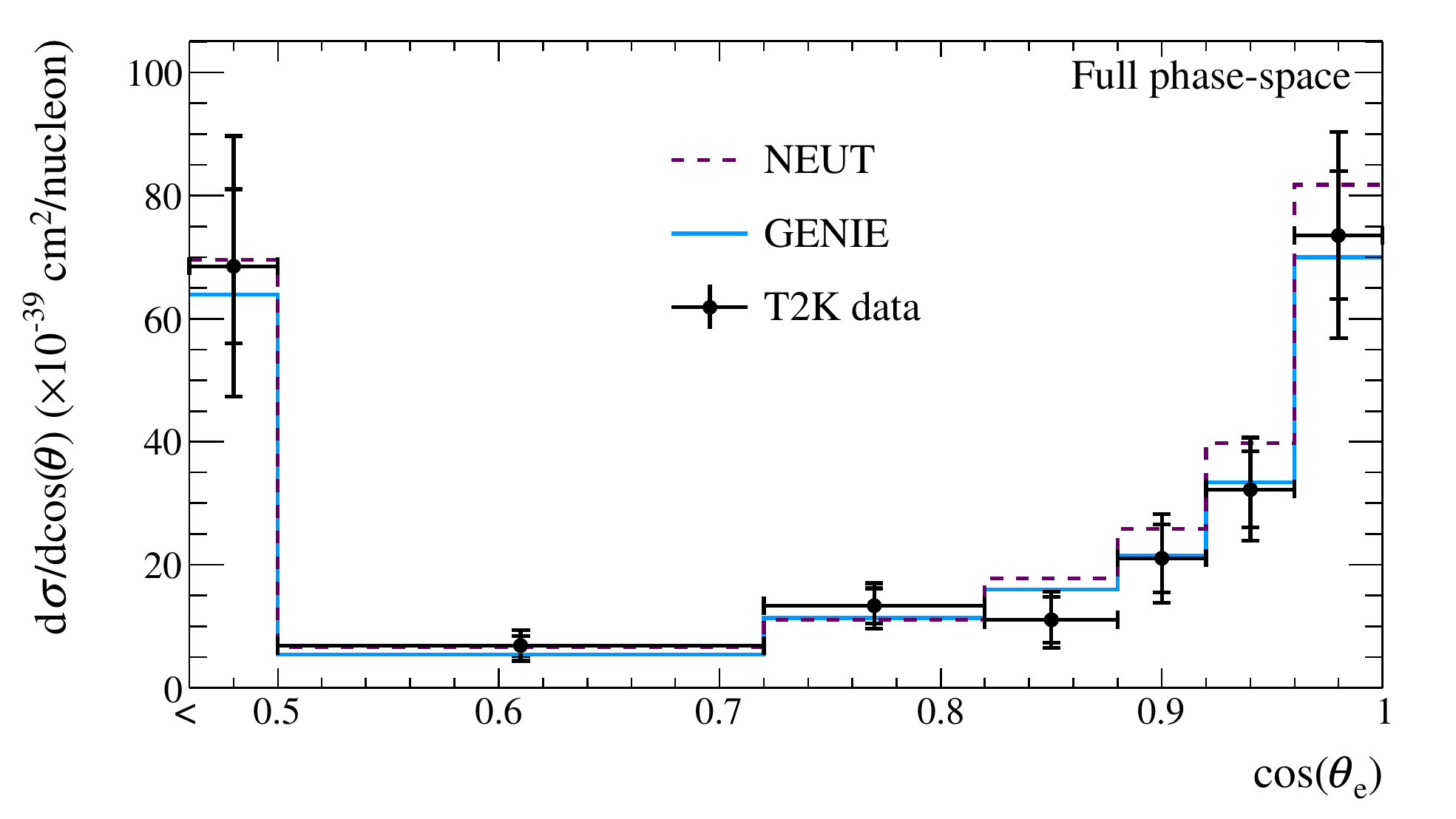}
      \caption{}
    \end{subfigure}
    \begin{subfigure}[b]{0.4\textwidth}
      \includegraphics[width=\textwidth]{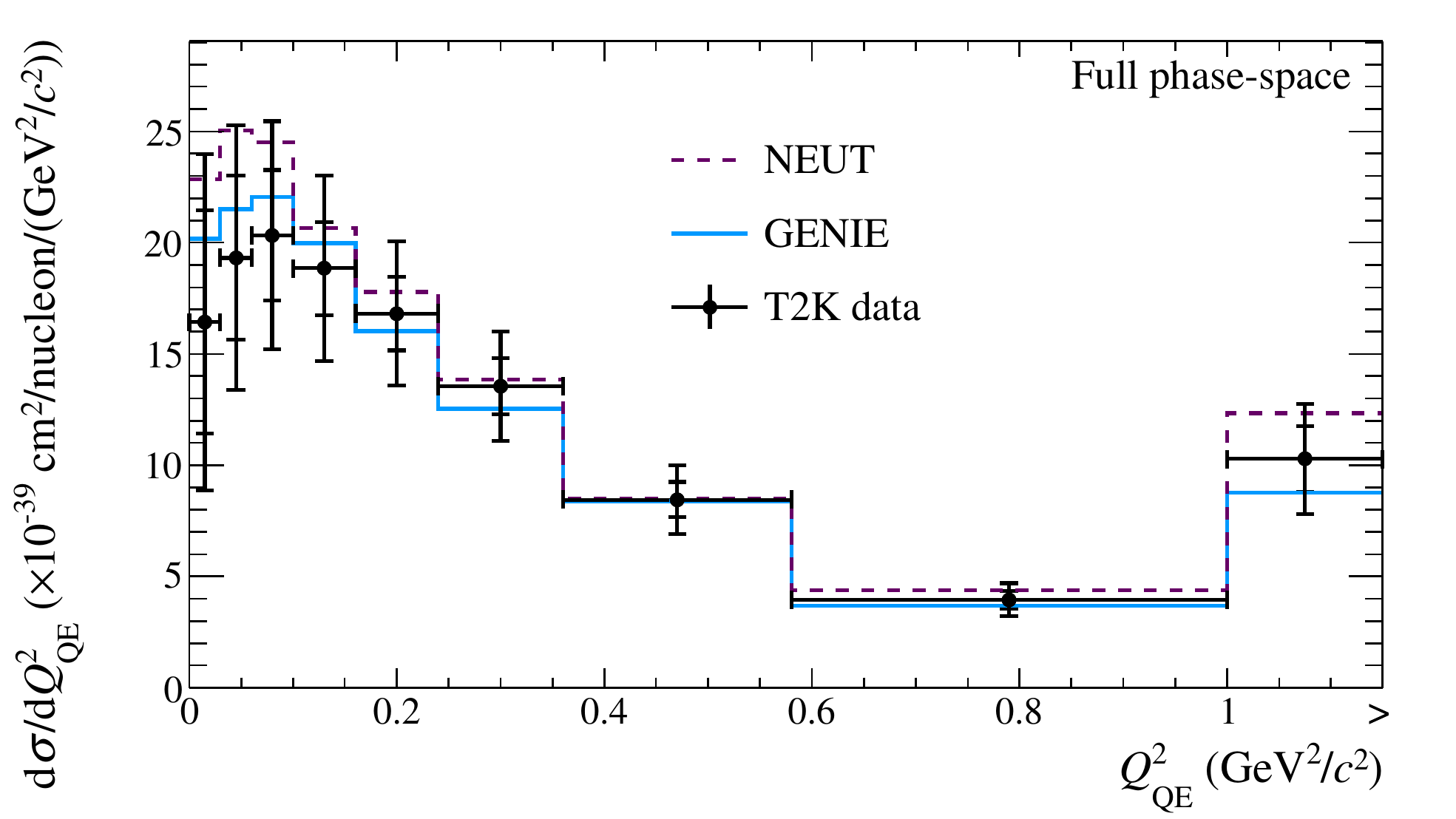}
      \caption{}
    \end{subfigure}
    \begin{subfigure}[b]{0.4\textwidth}
      \includegraphics[width=\textwidth]{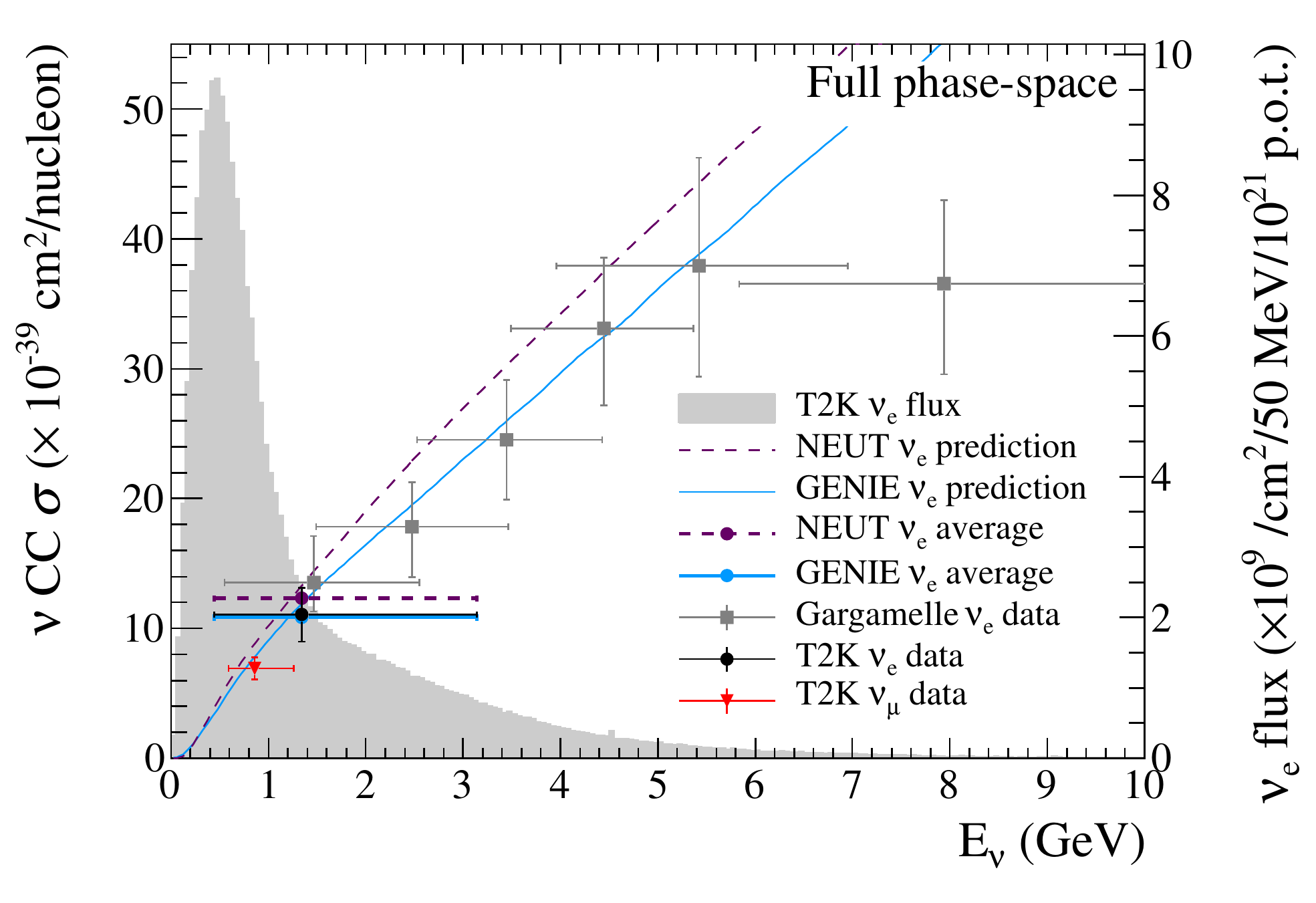}
      \caption{}
    \end{subfigure}
    
    \caption{$\nu_e$-CC cross sections on Carbon, integrated over the
      T2K beam flux. (a-c) show differential cross sections in
      electron momentum, interaction $Q^2$, and the cosine of the
      electron polar angle, respectively. (d) shows the total cross
      section.}
\label{fig:nuEXSec}
\end{center}
\end{figure}

Figure \ref{fig:nuEXSec}~(a-c) show the differential cross sections
(averaged over the T2K flux), as a function of electron momentum,
interaction $Q^2$, and the cosine of the electron polar angle,
respectively. Good agreement is seen between T2K and the generators
for the differential cross sections. Figure \ref{fig:nuEXSec}~(d)
shows the measured cross section, integrated over the T2K beam flux,
along with results from the Gargamelle bubble chamber experiment and
the NEUT and GENIE Monte Carlo interaction generators; again, good
agreement is seen with the generators, and also the Gargamelle data.

\subsection{$\nu_{\mu}$ CC-inclusive cross section on Carbon}

The ND280 detector has also measured the $\nu_{\mu}$ CC-inclusive
cross section on Carbon~\cite{Abe:2013jth}. The selection for this
measurement is essentially identical to that used for the analysis
constraining systematics for Super-K (see Section
\ref{Section:ND280}). All $\nu_\mu$-CC candidate events are included in
the selection.

\begin{figure}[!ht]
  \begin{center}
    \begin{subfigure}[b]{0.4\textwidth}
      \includegraphics[width=\textwidth]{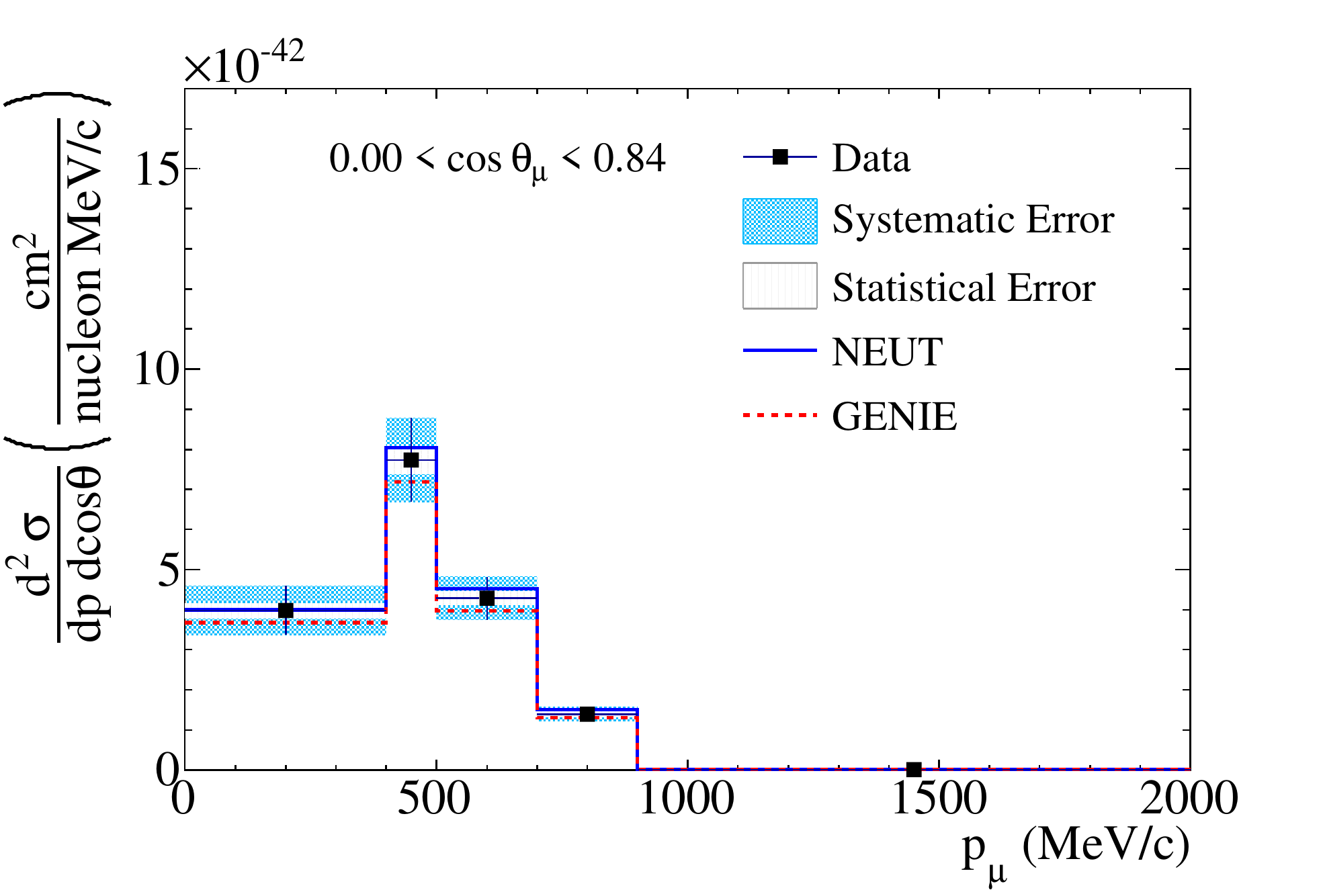}
      \caption{}
    \end{subfigure}
    \begin{subfigure}[b]{0.4\textwidth}
      \includegraphics[width=\textwidth]{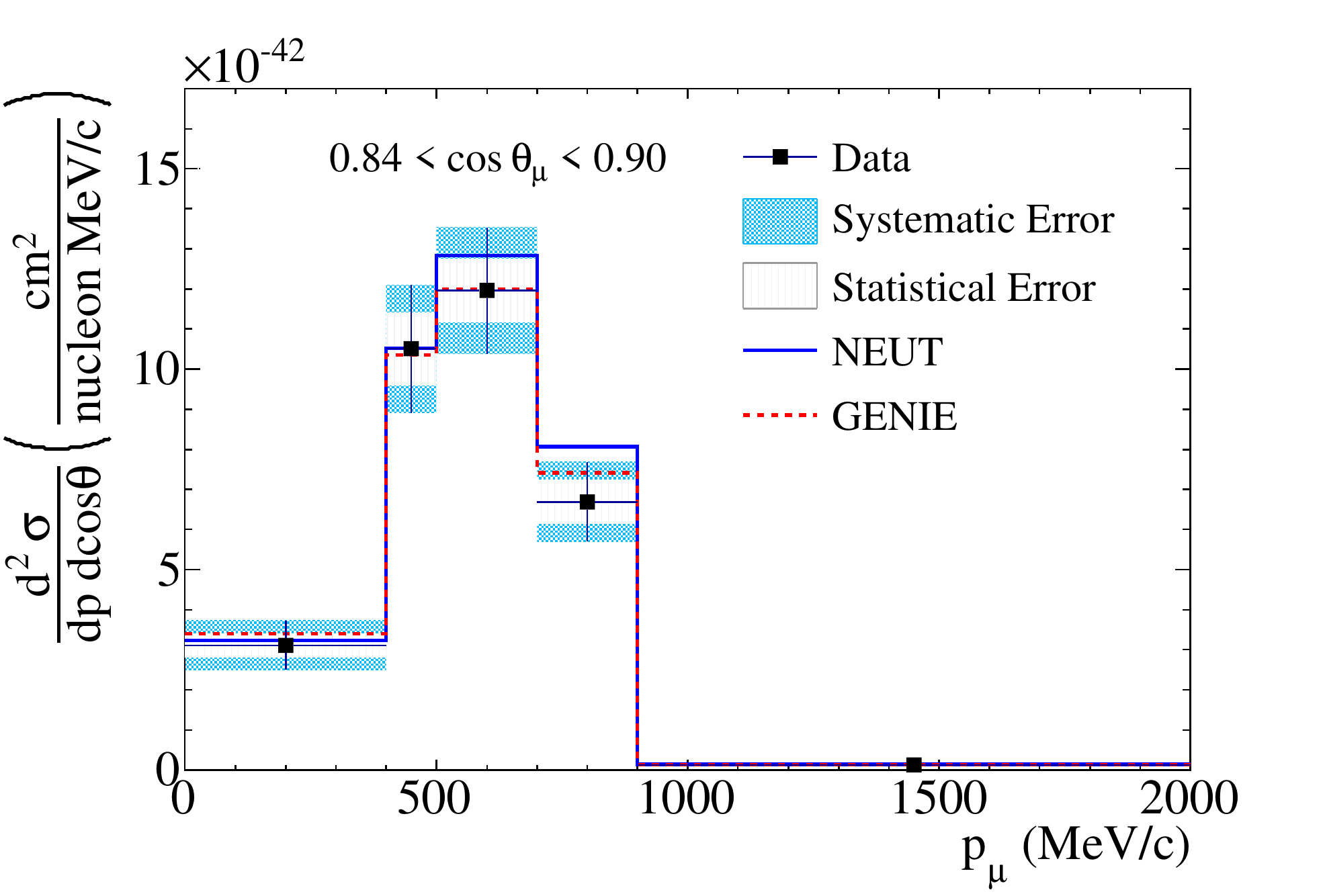}
      \caption{}
    \end{subfigure}
    \begin{subfigure}[b]{0.4\textwidth}
      \includegraphics[width=\textwidth]{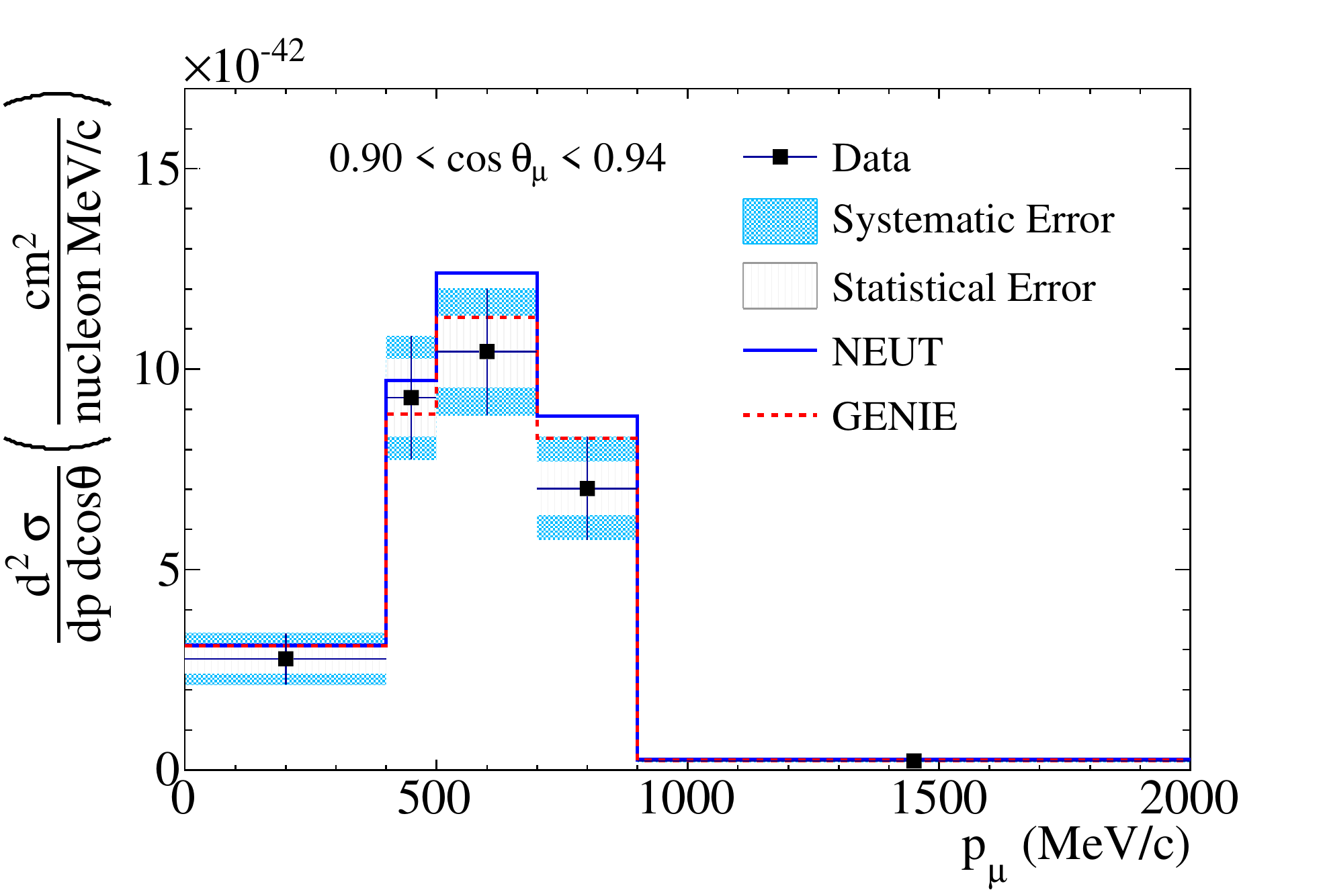}
      \caption{}
    \end{subfigure}
    \begin{subfigure}[b]{0.4\textwidth}
      \includegraphics[width=\textwidth]{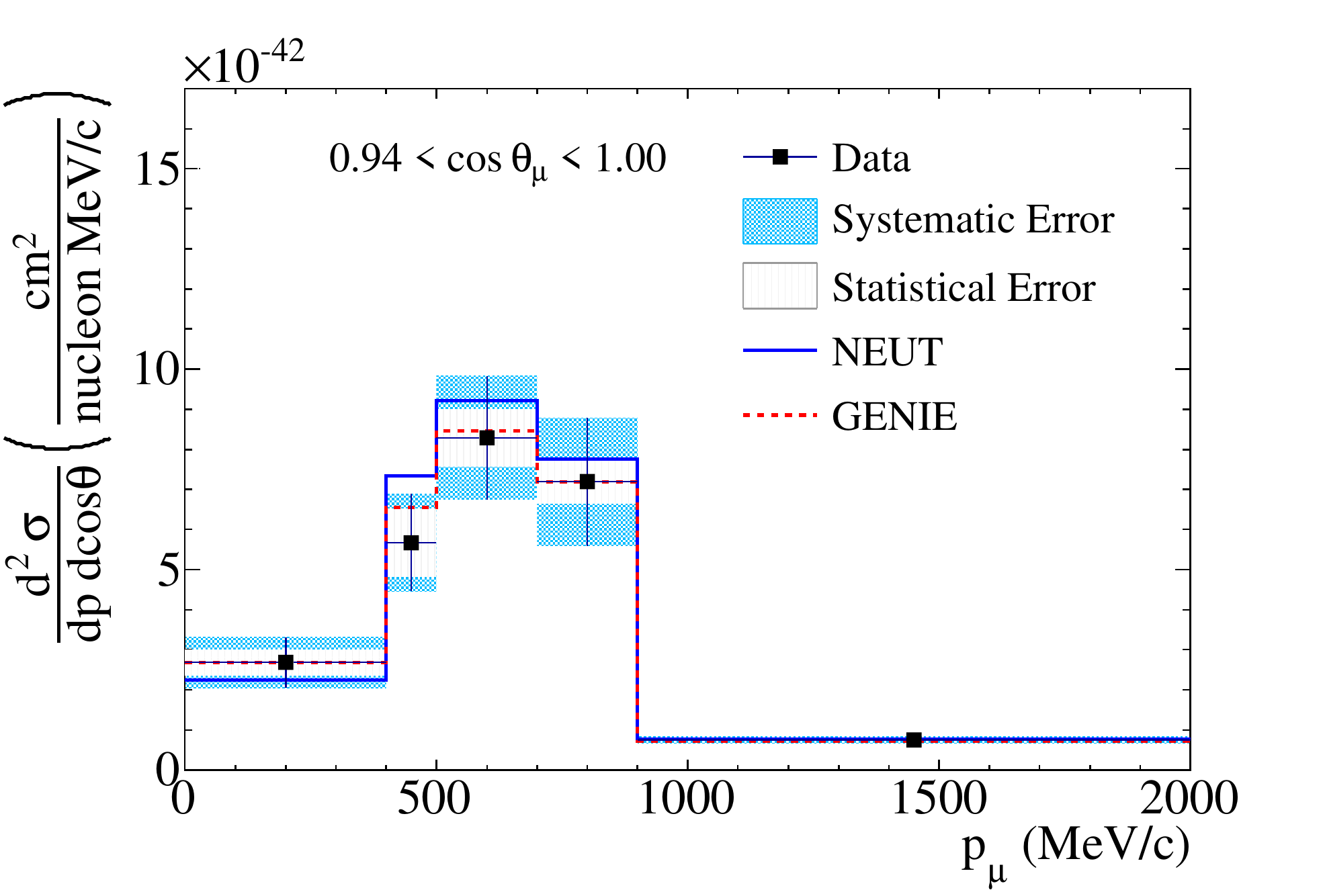}
      \caption{}
    \end{subfigure}
    
    \caption{$\nu_\mu$-CC cross sections on Carbon, integrated over
      the T2K beam flux. (a-d) each show the differential cross
      section in muon momentum, for a different range of muon polar
      angle.}
\label{fig:nuMuXSec}
\end{center}
\end{figure}

The events are binned in muon momentum $p_\mu$ and polar angle $\cos
\theta_\mu$, and differential cross sections are computed in these
values, integrated over the beam flux. These results are shown in
Figure \ref{fig:nuMuXSec} --- the results agree well with the NEUT and
GENIE generators.

\subsection{NCQE cross section from de-excitation gammas}

T2K has made the first ever measurement of neutral current
quasi-elastic (NCQE) neutrino scattering on
Oxygen~\cite{Abe:2014dyd}. This process, ${\nu_x + {}^{16}\textrm{O}
  \to \nu_x + p + {}^{15}\textrm{N}^*}$ (or $\nu_x + n +
{}^{15}\textrm{O}^*$), leaves the daughter nucleus in an excited
state, so the events can be tagged using gamma rays from nuclear
de-excitation. A beam with a low duty cycle, combined with precise
timing information, allows to reduce the non-beam background to a
negligible level, enabling a measurement which would be impossible
with atmospheric neutrinos.

After making standard Super-K cuts to remove non-beam background, and
vetoing on high-energy preceding events, events are selected which
have a visible energy between 4 and 20~MeV, and a large Cherenkov
angle. This Cherenkov angle cut helps to remove low-energy muons from
the event sample. After making these cuts, the dominant background is
from NC non-QE events.

\begin{figure}[!ht]
  \begin{center}
      \includegraphics[width=0.35\textwidth]{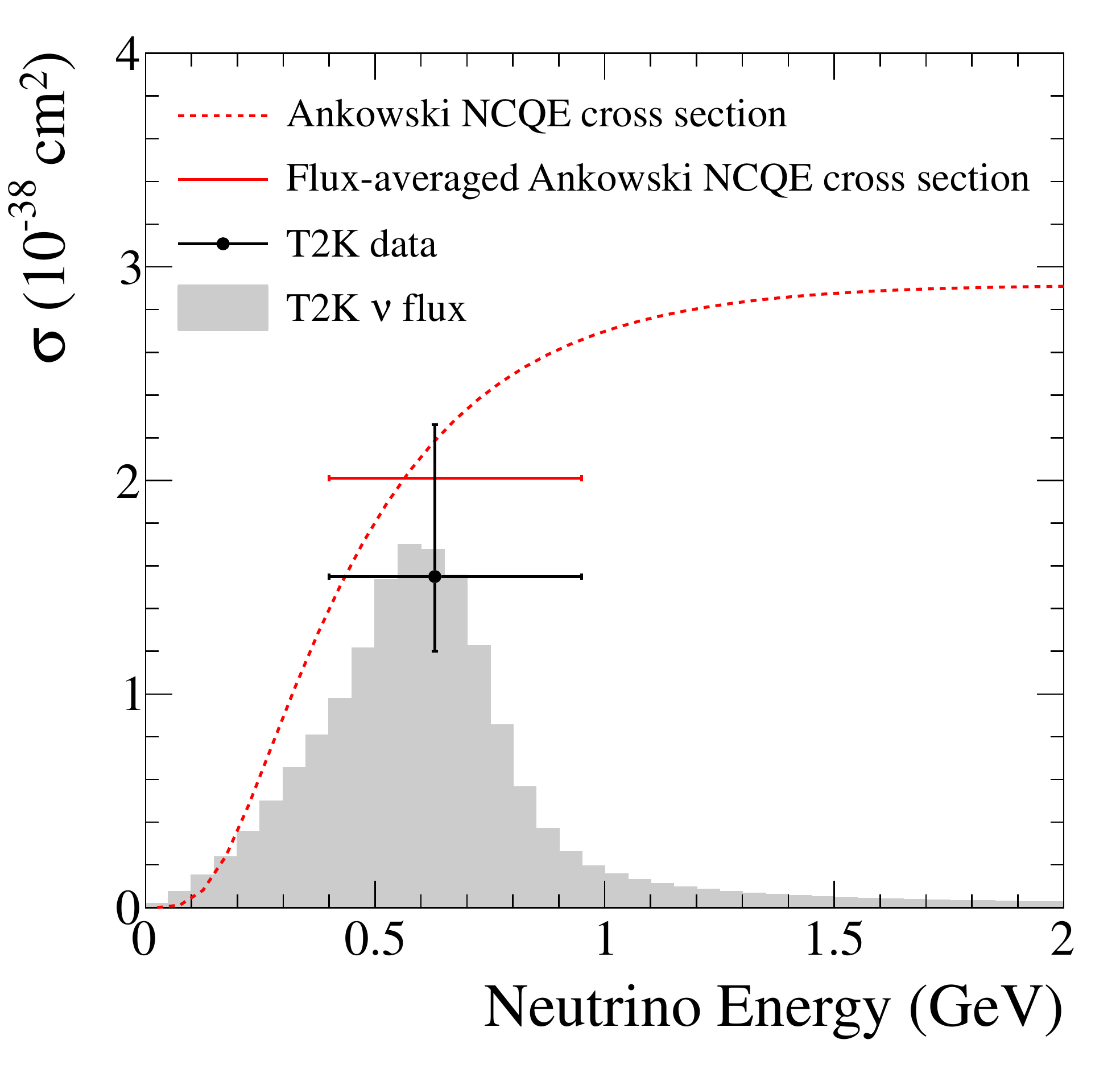}
    \caption{Measurement of the NCQE cross section for the T2K beam
      flux.}
\label{fig:NCQEXSec}
\end{center}
\end{figure}

Figure \ref{fig:NCQEXSec} shows the measured cross section, averaged
over the T2K beam flux. This is seen to be compatible with the
Ankowski NCQE model over the flux energy range.

\section{Summary}

The T2K experiment has published results showing exclusion of
$\theta_{13}=0$ of $7.3\sigma$ in the $\nu_e$ appearance channel, and
giving the world's most precise measurement of $\theta_{23}$. It has
also produced world-leading neutrino cross section results using both
near and far detectors.

\bigskip
\section{Acknowledgments}

I gratefully acknowledge the support of the STFC, which provides
funding for my participation in the T2K project.

\end{document}